\documentclass[aps, prd, twocolumn, 
showpacs, showkeys, 
print, 
groupedaddress,
nofootinbib
]{revtex4-2}

\usepackage[utf8]{inputenc}
\usepackage[T1]{fontenc}
\usepackage{lmodern}
\usepackage{microtype}
\usepackage[english]{babel}

\usepackage{amssymb,amsbsy,amsmath,amsfonts,amsthm}
\usepackage{yhmath} 

\usepackage{slashed}
\usepackage{bm}
\usepackage{mathptmx}

\usepackage[dvipsnames,table,xcdraw]{xcolor}
\definecolor{lightyellow}{RGB}{255,250,205}

\usepackage{graphicx}
\usepackage{epstopdf}

\usepackage{array}         
\usepackage{tabularx}
\usepackage{multirow}
\usepackage{longtable}

\usepackage{url}
\usepackage{orcidlink}
\usepackage{comment}

\usepackage{hyperref}  
\hypersetup{  
  colorlinks=true,
  linkcolor=blue,  
  citecolor=cyan,
  urlcolor=red,           
}

\begin{document}

\title{$S$-wave $KN$ scattering in a renormalizable chiral effective field theory}

\author{Xiu-Lei Ren\orcidlink{0000-0002-5138-7415}}
\email[]{xiulei.ren@sdu.edu.cn}
\affiliation{Shandong Provincial Key Laboratory of Nuclear Science, Nuclear Energy Technology and Comprehensive Utilization, 
\& School of Nuclear Science, Energy and Power Engineering, Shandong University, 
250061 Jinan, China}

\begin{abstract}
We investigate the $s$-wave $KN$ scattering up to next-to-leading order within a renormalizable framework of covariant chiral effective field theory. Using time-ordered perturbation theory, the scattering amplitude is obtained by treating the leading-order interaction non-perturbatively and including the higher-order corrections perturbatively via the subtractive renormalization. We demonstrate that the non-perturbative treatment is essential, at least at lowest order, in the SU(3) sector of $KN$ scattering. Our NLO study achieves a good description of the empirical $s$-wave phase shifts in the isospin $I=1$ channel. An analysis of the effective range expansion yields a negative effective range, consistent with some partial wave analyses but opposite in sign to earlier phenomenological summaries. For the $I=0$ counterpart, the $KN$ interaction is found to be rather weak and exhibits large uncertainties. Further low-energy $KN$ scattering experiments and lattice QCD simulations are needed  to better constrain both $s$-wave channels.
\end{abstract}

\date{\today}
\maketitle

\section{Introduction}

Meson-baryon scattering provides an ideal playground to deepen our understanding of Quantum Chromodynamics (QCD) in the non-perturbative regime, particularly due to the rich spectrum of baryonic resonances~\cite{Thiel:2022xtb}. 
Regarding the kaon-nucleon ($KN$) system, early interest was stimulated by the reported evidence for the 
$\Theta^+(1540)$ resonance, a candidate pentaquark state with strangeness $S=+1$, predicted in the chiral soliton model~\cite{Diakonov:1997mm} and later reported by the LEPS experiment in 2003~\cite{LEPS:2003wug}. 
However, it was disfavored by the subsequent precise experiments, as summarized in Ref.~\cite{Liu:2014yva},
and theoretical analysis of LEPS data suggests that the $\Theta^+(1540)$ may not be a genuine resonance~\cite{MartinezTorres:2010zzb}. 
Apart from this argument, the $KN$ interaction itself exhibits several interesting features. 
At low energies, the $KN$ interaction is relatively weak and unaffected by nearby inelastic channels.
This feature allows one to investigate the effective potential and nuclear structure by using the low-energy kaon beams scattering off nuclei, 
in which the kaon can penetrate deeply into the nuclear interior~\cite{Dover:1982zh,Friedman:2007zza}. 
Accurate knowledge of the low-energy $KN$ scattering is also essential for studies of the partial restoration of chiral symmetry and the strange quark condensate in the nuclear medium~\cite{Aoki:2017hel,Iizawa:2023xsi}. 
Moreover, the  elastic $KN$  scattering provides a clean channel for exploring the SU(3) flavor dynamics in QCD. 

Experimentally, the elastic $KN$ scattering was measured several decades ago~\cite{Goldhaber:1962zz,Stenger:1964vej,Bowen:1970azd,Cameron:1974xx,Burnstein:1974ax,Glasser:1977xs}, with the incident kaon momentum ($P_\mathrm{lab}$) below $400$ MeV, as summarized in Refs.~\cite{Dover:1982zh,Dumbrajs:1983jd}. 
However, the lowest available $P_\mathrm{lab}$ value is around $130$ MeV, which is not sufficiently close to the $KN$ threshold.  
As a result, the low-energy parameters, such as the scattering length and effective range, have been extracted via the energy-dependent partial wave analyses (PWAs)~\cite{Martin:1975gs,Nakajima:1982tk,Hashimoto:1984th,Hyslop:1992cs,Gibbs:2006ab}. 
On the theoretical side, the $KN$ interaction has been investigated 
using phenomenological approaches~\cite{Veit:1984sf,Buettgen:1990yw,Hoffmann:1995ie,Hadjimichef:2002xe,Silvestre-Brac:1995lkn,Silvestre-Brac:1997azu,Lemaire:2001sr,Lemaire:2003et,Lemaire:2003py,Huang:2004sj,Huang:2004ke,Huang:2005hy,Polinder:2005sn,Wu:2007fc}, such as the J\"ulich meson-exchange model~\cite{Buettgen:1990yw,Hoffmann:1995ie,Hadjimichef:2002xe} and the chiral SU(3) quark model~\cite{Huang:2004sj,Huang:2004ke,Huang:2005hy}. 
In addition, the $KN$ scattering length and amplitude have been studied within chiral perturbation theory~\cite{Kaiser:2001hr,Liu:2006xja,Mai:2009ce,Huang:2015ghe,Lu:2018zof,Ou-Yang:2025oue} and its unitarized extension~\cite{Khemchandani:2014ria,Aoki:2018wug}.  
These studies, particularly for the $s$-wave $KN$ scattering, indicate a repulsive interaction in the $I = 1$ channel at the qualitative level, 
with the scattering length typically in the range $[-0.33, -0.28]$ fm~\cite{Dover:1982zh}. In contrast, the $s$-wave $I = 0$ channel remains poorly constrained due to the large uncertainties in the low-momentum kaon data.
Motivated by this situation, the HAL QCD collaboration has recently reported their lattice simulations of $s$-wave $KN$ scattering at the physical pion mass~\cite{Murakami:2025owk}. 
Although this study confirms repulsion in the $I = 1$ channel, the extracted scattering length is smaller,
differing from the summary result~\cite{Dover:1982zh} by about  $2\sim 3\,\sigma$. 
In the $I = 0$ channel, the lattice calculation indicates only a very weak interaction.

Motivated by the remaining uncertainties in the low-energy $KN$ interaction, 
and the current tension between phenomenological analyses and the recent HAL QCD result,
in this work, we investigate the $KN$ interaction by using time-ordered perturbation theory (TOPT) within a renormalizable framework of covariant chiral effective field theory (ChEFT), as proposed in Refs.~\cite{Ren:2020wid,Epelbaum:2020maf} for meson-baryon systems~\footnote{This scheme has been applied to the studies of chiral nucleon-nucleon and hyperon-nucleon interactions in Refs.~\cite{Baru:2019ndr,Ren:2019qow,Ren:2022glg,Ren:2025pvy}.}. 
This framework has been successfully applied to study the $\pi N$ scattering in the SU(2) sector~\cite{Ren:2020wid}, 
and to the strangeness $S=-1$ meson-baryon interactions, i.e. the $\bar{K}N$ scattering with coupled channels, in the SU(3) sector up to leading order (LO)~\cite{Ren:2021yxc,Ren:2024frr}. 
In contrast to these systems, low-energy $KN$ scattering, as a single channel case, is free from nearby resonance contributions. 
This simplicity allows for a systematic extension of the renormalizable ChEFT framework to next-to-leading order (NLO), which we perform here for the first time. 
Then, the $KN$ system provides a typical playground to investigate the convergence properties and predictive power of the renormalizable ChEFT in the SU(3) sector, without complications from the explicit resonance contribution, such as the $\Delta$ isobar that plays an essential role in $\pi N$ scattering. It is worth emphasizing that the convergence of chiral EFT in the SU(3) baryon sector faces a general challenge due to the less pronounced scale separation compared to the SU(2) case. The present work could be regarded as an exploratory step, and a systematic order-by-order convergence study will require extending the current calculation to higher orders, e.g. at least next-to-next-to-leading order (NNLO).

The paper is organized as follows. In Sec.~\ref{sec2}, we formulate the chiral effective potential of $KN$ scattering up to NLO, and obtain the renormalized $T$-matrix by solving the scattering integral equations and utilizing a subtractive renormalization scheme. Numerical evaluation of the $s$-wave phase shifts and effective range expansion at LO and NLO are presented in Sec.~\ref{sec3}. Finally, we summarize our findings in Sec.~\ref{sec4}.

\section{Theoretical Framework} \label{sec2}

\subsection{$KN$ scattering amplitude}
The on-shell amplitude of the $KN$ scattering process, $K(q_1)+N(p_1) \to K(q_2) + N(p_2)$, 
 can be parametrized as
 \begin{equation}\label{eq:TKN}
 \begin{aligned}
 	 T_{KN} &= \bar{u}_{N}(p_2,s_2) \left[A+\frac{1}{2}(\slashed q_1 + \slashed q_2) B\right] u_{N}(p_1,s_1) \\
  &= \bar{u}_{N}(p_2,s_2) \left[D+\frac{i}{2m_{N}}\sigma_{\mu\nu}\, q_{2}^{\mu} q_{1}^{\nu}\, B\right] u_{N}(p_1,s_1)\, ,
 \end{aligned}	
 \end{equation}
where $D=A+(s-u)B/(4m_{N})$ with the Mandelstam variables: $s=(p_1+q_1)^2$, $t=(p_1-p_2)^2$, and $u=(p_1-q_2)^2$. 
 The Dirac spinor $u_N(p,s)$ is normalized as
\begin{equation}\label{Eq:diracspinor}
u_N(p,s)=\sqrt{\frac{\omega_N(\bm{p})+m_N}{2 m_N}}\left(\begin{array}{c}
1 \\
\frac{\bm{\sigma} \cdot \bm{p}}{\omega_N(\bm{p})+m_N}
\end{array}\right)\chi_s,
\end{equation}
where $\chi_s$ is a two-component spinor with spin $s$, and  $\omega_N(\bm{p})=\sqrt{\bm{p}^2+m_N^2}$ is the nucleon energy. One can reduce the $KN$ scattering amplitude (Eq.~\eqref{eq:TKN}) in terms of the Pauli matrix 
\begin{equation}
	T_{KN} = T_{KN}^{c} + i\, T_{KN}^{so}\, \bm{\sigma}\cdot (\bm{q}_2\times\bm{q}_1),
\end{equation}
with the non-spin-flip amplitude $T_{KN}^{c}$ and the spin-flip amplitude $T_{KN}^{so}$. 

For convenience in later calculations, we express the scattering amplitude in the center-of-mass (CM) frame with the following four-momentum assignments 
\begin{equation}
\begin{aligned}
q_{1}^\mu &=\left(\omega_{K}(\bm{p}), \bm{p}\right),\quad p_{1}^\mu
=\left(\omega_{N}(\bm{p}),-\bm{p}\right),\quad  \\
q_{2}^\mu &=\left(\omega_{K}(\bm{p}'), \bm{p}'\right),\quad p_{2}^\mu
=\left(\omega_{N}(\bm{p}'),-\bm{p}'\right),
\end{aligned}
\end{equation}
where the relative incoming and outgoing momenta $\bm{p}$, $\bm{p}'$ are introduced.

\subsection{Chiral effective Lagrangian}
The lowest order chiral effective Lagrangian describing the interaction between the pseudoscalar meson octet ($\pi$, $K$, $\eta$) 
and the lowest-lying baryon octet ($N$, $\Lambda$, $\Sigma$, $\Xi$) is given by~\cite{Ren:2020wid} 
\begin{equation}\label{Eq:LagLO}
\begin{aligned}
	\mathcal{L}_\mathrm{LO} &= \frac{F_0^2}{4}\, \left\langle u_\mu
u^\mu +\chi_+ \right\rangle  \\
& \quad + \left\langle \bar{B} \left( i\gamma_\mu \partial^\mu -m \right)  B \right\rangle + \frac{D/F}{2}
 \left\langle \bar{B} \gamma_\mu \gamma_5 [u^\mu,B]_{\pm}\right\rangle   \nonumber\\
&\quad - \frac{1}{4} \, \left\langle  V_{\mu\nu}   V^{\mu\nu} - 2 \mathring{M}_V^2 \,\left( V_{\mu} - i \, \Gamma_\mu/g \right)^2 \right\rangle  \\
&\quad +  g \, \left\langle \bar{B} \gamma_\mu [V^\mu,B]\right\rangle,
\end{aligned}
\end{equation}
where $\langle \ldots \rangle$ denotes the trace  in the flavor space, and $\chi_{+} = u^{\dagger} \chi u^{\dagger} + u \chi^{\dagger} u$,
$u_{\mu}= i (u^{\dagger} \partial_{\mu}u - u\partial_{\mu}u^{\dagger})$, $\Gamma_{\mu} = \frac{1}{2}\left(u^{\dagger} \partial_{\mu} u+u \partial_{\mu} u^{\dagger}\right)$ with  $u=\exp \left( i\, \Phi /(2F_{0}) \right)$ and $\chi=2 B_{0}\,\mathrm{diag}(m_l, m_l, m_s)$. The axial vector couplings of baryons to the mesons ($D$, $F$) are taken as $D=0.760$ and $F=0.507$, satisfying the SU(6) relation $F = 2/3D$ and $D+F=g_A=1.267$. $F_0$ denotes the chiral limit of pseudoscalar meson decay constant. {In the case of $KN$ scattering, we can take $F_0=F_K=110.03$ MeV.} Note that vector meson octet ($\rho$, $\omega$, $K^*$, $\phi$) are explicitly included in the second line of Eq.~\eqref{Eq:LagLO} with $V_{\mu \nu} = \partial_{\mu} V_{\nu}-\partial_{\nu} V_{\mu} - ig[V_\mu,V_\nu]$, the coupling $g=\mathring{M}_V/(\sqrt{2}F_0)$ determined by the KSFR relation, and the chiral limit value of vector meson mass $\mathring{M}_V$. 

Up to next-to-leading order, the effective Lagrangian for the $s$-wave meson-baryon scattering can be expressed as~\cite{Borasoy:2002mt}
\begin{equation}
\begin{aligned}
  \mathcal{L}_\mathrm{NLO} & =b_0 \langle \chi_+ \rangle \langle\bar{B}B\rangle
  +b_{D}\langle\bar{B}\{\chi_+, B\}\rangle  + b_{F}\langle \bar{B}[\chi_+, B] \rangle \\
  &\quad + d_1 \bigl\langle \bar{B} \{ u_\mu, [u^\mu, B]\} \bigr\rangle 
     + d_2 \bigl\langle \bar{B} \bigl[ u_\mu, [u^\mu, B]\bigr] \bigr\rangle \\
     &\quad 
     + d_3 \langle\bar{B}u_\mu\rangle \langle u^\mu B\rangle 
     + d_4 \langle \bar{B}B\rangle \langle u^\mu u_\mu\rangle, 
\end{aligned}
\end{equation}
where the first line is the  explicit chiral symmetry breaking terms with the low-energy constants (LECs) $b_{0}$, $b_D$, and $b_F$. The latter four terms $d_{1,...,4}$ with the two derivatives can also contribute to the $s$-wave meson-baryon potential.

\begin{figure}[t]
\centering 
  \includegraphics[width=0.48\textwidth]{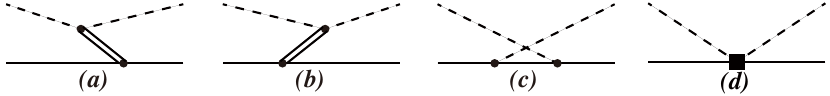}
  \caption{Time-ordered diagrams for $KN$ scattering up to NLO. The dashed, solid and double-solid lines correspond to kaon, octet baryons and vector mesons, respectively. The dots (boxes) denote the $\mathcal{O}(p^1)$ ($\mathcal{O}(p^2)$) vertices.}
    \label{Fig:diagramsNLO}
\end{figure}

\subsection{$KN$ interaction up to NLO }
In this section, we present the $KN$ interaction up to NLO in the TOPT framework, where the Dirac spinor~\eqref{Eq:diracspinor} is perturbatively expanded as follows: 
\begin{equation}\label{Eq:expanddirac}
\begin{aligned}
u_N(p,s) &=  u_0 + u_1 + u_2 + \cdots \\
&= \left[ \begin{pmatrix}
   1\\
   0	
 \end{pmatrix} 
 + \frac{1}{2m_N}\begin{pmatrix}
   0\\
   \bm{\sigma}\cdot\bm{p}	
 \end{pmatrix}
 + \frac{1}{8m_N^2} 
 \begin{pmatrix}
   \bm{p}^2	\\
   0
 \end{pmatrix}
 + ... \right]\, \chi_s.
\end{aligned}
\end{equation}
In this way, we can employ the standard power counting to organize the effective potential up to $\mathcal{O}(p^2)$.

At leading order, the $KN$ scattering amplitude for the process $K(q_1)+ N(p_1) \to K(q_2) +N(p_2)$ is given by the time-ordered diagrams Fig.~\ref{Fig:diagramsNLO}(a-c).  
According to the diagrammatic rules of TOPT~\cite{Baru:2019ndr,Ren:2024bkz}, one can obtain the LO potential:
\begin{equation}
	V_\mathrm{LO} = V_\mathrm{VME} + V_\mathrm{CB},
\end{equation} 
where the Born term does not contribute. 
Instead of the Weinberg-Tomozawa contact term, we have the vector-meson-exchange (VME) contribution, which can be expressed as 
\begin{equation}
\begin{aligned}
V_\mathrm{VME} &=  -\frac{1}{32 F_K^2} \sum\limits_{V=\rho, \omega,\phi} C_{KN,KN}^V \frac{M_V^2\left[\omega_{K}(\bm{q}_1)+\omega_{K}(\bm{q}_2)\right] }{\omega_V(\bm{q}_1-\bm{q}_2)} \\
  &\quad \times \Biggl[\frac{1}{E-\omega_{N}(\bm{p}_1)-\omega_V(\bm{q}_1-\bm{q}_2)-\omega_{K}(\bm{q}_2)}  \\
  &\qquad + \frac{1}{E-\omega_{N}(\bm{p}_2)-\omega_V(\bm{q}_1-\bm{q}_2)-\omega_{K}(\bm{q}_1)}  \Biggr],
\end{aligned}
\end{equation}
where the particle energy $\omega_{K,N,V}(\bm{p})$ is defined as $\sqrt{\bm{p}^2 + M_{K,N,V}^2}$, and $E$ is the total energy of $KN$ system. 
The coefficients $C_{KN,KN}^V$ for vector mesons $V=\rho$, $\omega$, $\phi$ take the following values: 
 $\{C_{KN,KN}^{\rho},\, C_{KN,KN}^{\omega}, \, C_{KN,KN}^{\phi}\}=\{-6,\, 2,\, 4\}$ in the isospin $I=0$ sector, 
 and in the $I=1$ sector, they become $\{2,\,2,\,4\}$. 
In practice, we use the physical vector-meson masses $M_{\rho,\omega,\phi}$ instead of the chiral limit value $\mathring{M}_V$, and the difference due to this approximation is of higher order~\cite{Ren:2020wid,Ren:2021yxc}.
It is worth mentioning that the VME contribution in the above equation is consistent with  Weinberg-Tomozawa term  in the heavy vector meson mass limit. 

As to the crossed-Born term, its expression is given as 
\begin{equation}
\begin{aligned}
	 V_\mathrm{CB} & =  \frac{1}{4F_K^2}\sum\limits_{B=\Lambda,\Sigma} \tilde{C}_{KN,KN}^{B} \frac{m_{B}}{\omega_{B}(\bm{K})} \\
	 &\quad \times \frac{(\bm{\sigma}\cdot\bm{q}_1)(\bm{\sigma}\cdot\bm{q}_2)}{E-\omega_{K}(\bm{q}_1)-\omega_{K}(\bm{q}_2)-\omega_{B}(\bm{K})}\,,
\end{aligned}
\end{equation}
with $K^\mu = p_1^\mu -q_2^\mu = p_2^\mu - q_1^\mu$.
The coefficients 
$\tilde{C}_{KN,KN}^{B}$ for $B=\Lambda$, $\Sigma$ are expressed in terms of $D$ and $F$ as: 
$\tilde{C}_{KN,KN}^{\Lambda} = -\frac{1}{3}(D+3F)^2$, $\tilde{C}_{KN,KN}^{\Sigma} = 3(D-F)^2$ in the isospin $I=0$ sector, and  
$\tilde{C}_{KN,KN}^{\Lambda} = (D+3F)^2/3$, $\tilde{C}_{KN,KN}^{\Sigma} = (D-F)^2$ in the isospin $I=1$ sector.

Up to NLO, the $KN$ effective potential receives contributions from two sources:
\begin{equation}
	V_\mathrm{NLO} = V_\mathrm{CT}^{(2)} + V_\mathrm{LO}^{(2)},
\end{equation}
where $V_\mathrm{CT}^{(2)}$ is the contact term (Fig.~\ref{Fig:diagramsNLO}d) with the second order vertex, 
and $V_\mathrm{LO}^{(2)}$ represents the NLO correction to the VME potential and crossed-Born term, 
obtained from higher-order terms in the spinor decomposition of Eq.~\eqref{Eq:diracspinor}. 

In the isospin basis,  the NLO contact terms are 
\begin{equation}
\begin{aligned}
	V_\mathrm{CT}^{(2),I=0} & = \frac{4(b_0-b_F)}{F_K^2} M_K^2 + \frac{2(2d_1+d_3-2d_4) }{F_K^2}  \bm{q}_1\cdot \bm{q}_2,\\
	V_\mathrm{CT}^{(2),I=1} & =\frac{4(b_0+b_D)}{F_K^2} M_K^2  - \frac{2(2d_2+d_3+2d_4)}{F_K^2} \bm{q}_1\cdot \bm{q}_2,
\end{aligned}
\end{equation}
where the combinations of LECs are given as  
\begin{equation}~\label{Eq:4LECs}
\begin{aligned}
	b^{I=0} & = 4(b_0-b_F),\quad 
	d^{I=0} = 2(2d_1+d_3-2d_4) ,\\
	b^{I=1} & = 4(b_0+b_D),\quad 
	d^{I=1} = 2(2d_2+d_3+2d_4),
\end{aligned}
\end{equation} 
which results in a total of $4$ unknown parameters for the $KN$ potential at NLO. 

As to the NLO correction from the VME and crossed-Born terms, $V_\mathrm{LO}^{(2)}=V_\mathrm{VME}^{(2)}+V_\mathrm{CB}^{(2)}$, 
we have the following expressions: 
\begin{widetext}
\begin{equation}
\begin{aligned}
V_\mathrm{VME}^{(2)} &=  -\frac{1}{32 F_K^2} \sum\limits_{V=\rho, \omega,\phi} C_{KN,KN}^V \frac{M_V^2}{\omega_V(\bm{q}_1-\bm{q}_2)} 
   \frac{1}{2m_{N}}\Bigl[ 2 \bigl( \bm{\sigma}\cdot \bm{q}_2\, \bm{\sigma}\cdot\bm{q}_1 \bigr) +\bm{q}_1^{\, 2}+  \bm{q}_2^{\, 2} \Bigr] \\
  &\quad \times \Biggl[\frac{1}{E-\omega_{N}(\bm{p}_1)-\omega_V(\bm{q}_1-\bm{q}_2)-\omega_{K}(\bm{q}_2)}  + \frac{1}{E-\omega_{N}(\bm{p}_2)-\omega_V(\bm{q}_1-\bm{q}_2)-\omega_{K}(\bm{q}_1)}  \Biggr],\\
	V_\mathrm{CB}^{(2)} &= \frac{1}{4F_K^2}\sum\limits_{B=\Lambda,\Sigma} \tilde{C}_{KN}^{B} \frac{m_{B}}{\omega_{B}(\bm{K})} \frac{1}{E-\omega_{K}(\bm{q_1})-\omega_{K}(\bm{q_2})-\omega_{B}(\bm{K})} \\
   &\quad \times \Biggl\{  \frac{m_{N}-m_B}{2m_B}\bigl( q_1^0 q_2^0 + 2 \bm{p}\cdot\bm{p}'  \bigr) + \frac{q_1^0}{4m_{N}m_B} \biggl[ 4m_{N}\bm{p}\cdot\bm{p}'   + (m_{N}+m_B){\bm{p}'}^2 - 2 m_{N} M_{K}^2 \biggr] \\
	& \qquad +\frac{q_2^0}{4m_{N} m_B} (3m_{N}+m_B) {\bm{p}}^2  \\
	&\qquad +\Biggl[ - \frac{m_{N}-m_B}{2m_B} - \frac{q_1^0}{4m_{N}m_B} (m_{N}+m_B)  +\frac{q_2^0}{4m_{N} m_B} (m_{N}-m_B)  \Biggl] \, (\bm{\sigma}\cdot {\bm{q}}_2)(\bm{\sigma}\cdot {\bm{q}}_1)  \Biggr\}.
\end{aligned}	
\end{equation}
\end{widetext}

Finally, we rewrite the NLO potential as central and spin-orbital parts,
\begin{equation}
\begin{aligned}
	V_{KN} &= V_\mathrm{LO} + V_\mathrm{NLO} \\
	&= V_{KN}^{c} + i \, V_{KN}^{so}\,\bm{\sigma}\cdot (\bm{q}_2\times \bm{q}_1),
\end{aligned} 
\end{equation}
which is convenient for us to perform the partial wave projection and obtain the $s$-wave potential via 
\begin{equation}
\begin{aligned}
	\langle L_{\pm} J| V_{KN} |L_{\pm} J\rangle
    & = 2\pi \int_{-1}^1 dz \, \biggl[ V^c_{KN} P_{L_\pm}(z) \\
      &\quad  + p^2 V^{so}_{KN} P_{L_{\pm} \pm 1 }(z) 
  - z p^2 V^{so}_{KN} P_{L_{\pm}}(z) \biggr],
\end{aligned}
\end{equation} 
with $z=\cos\theta$, $\theta$ as the cross angle between $\bm{p}$ and $\bm{p}'$, and $P_L(z)$ the conventional Legendre polynomial. 
The partial wave states are denoted as $|LJ\rangle$ with total angular momentum  $J$  and orbital angular momentum $L_\pm =J\mp 1/2$.

\subsection{$T$-matrix up to NLO}

The $T$-matrix of $KN$ scattering up to and including NLO is given by 
\begin{equation}
	T_{KN} = T_\mathrm{LO} + T_\mathrm{NLO}, 
\end{equation} 
where the LO contribution is obtained by solving the scattering integral equation 
\begin{equation}\label{Eq:TLO}
T_\mathrm{LO}(\bm{p}',\bm{p}) = V_\mathrm{LO}(\bm{p}',\bm{p})   +
\int\frac{d ^3\bm{k}}{(2 \pi)^3} V_\mathrm{LO}(\bm{p}',\bm{k})\, G(\bm{k}) \, T_\mathrm{LO}(\bm{k},\bm{p}),
\end{equation} 
while the NLO correction of $T$-matrix comes from treating the higher-order potential perturbatively,
\begin{equation}\label{Eq:TNLO}
\begin{aligned}
T_\mathrm{NLO}(\bm{p}',\bm{p})  & =   V_\mathrm{NLO}(\bm{p}',\bm{p}) \\
&\quad + \int\frac{d^3\bm{k}}{(2\pi)^3}T_\mathrm{LO}(\bm{p}',\bm{k}) G(\bm{k}) V_\mathrm{NLO}(\bm{k},\bm{p}) \\
&\quad + \int\frac{d^3 \bm{k}}{(2\pi)^3} V_\mathrm{NLO}(\bm{p}',\bm{k}) G(\bm{k}) T_\mathrm{LO}(\bm{k},\bm{p}) \\
&\quad + \int\int\frac{d^3 \bm{k}_1}{(2\pi)^3}\frac{d^3 \bm{k}_2}{(2\pi)^3}  T_\mathrm{LO}(\bm{p}',\bm{k}_1)  G(\bm{k}_1)  \\
&\qquad \times V_\mathrm{NLO}(\bm{k}_1,\bm{k}_2)  G(\bm{k}_2) T_\mathrm{LO}(\bm{k}_2,\bm{p}).
\end{aligned}
\end{equation}
The $KN$ two-body Green function can be written as 
\begin{equation}\label{Eq:GKN}
G(\bm{k})= \frac{1}{2\, \omega_K(\bm{k}) \omega_N(\bm{k})}\, \frac{m_N}{E - \omega_K(\bm{k}) -\omega_N(\bm{k})+i \epsilon} \,,
\end{equation} 
according to the time-ordered diagrammatic rules in TOPT framework. 

To obtain the renormalizable $T$-matrix of $KN$ scattering, we employ the subtractive renormalization scheme and 
replace the Green function in Eqs.~(\ref{Eq:TLO},~\ref{Eq:TNLO}) with the subtracted one $G_\mathrm{sub}$, defined as 
\begin{equation}
	G_\mathrm{sub}(E) = G(E) - G_e,
\end{equation}
with 
\begin{equation}
G_e=\left.\sum_{i=0}^N \frac{1}{i!}\left(E+E_\mu\right)^i \frac{d^i G(E)}{(d E)^i}\right|_{E=-E_\mu} .
\end{equation}
That is, we expand $G(E)$ at a fixed value $E=-E_\mu$ and subtract first several terms. The number of the subtracted terms depends on the UV behaviour of the singular potential. 
In this way, all divergences in loop diagrams are subtracted, which corresponds to including the contributions of an infinite  number of $KN$ counter-terms (details can be found in Refs.~\cite{Epelbaum:2020maf,Ren:2020wid}). This scheme can also remove the power-counting breaking terms originated from the iteration of the one-vector-meson exchange contributions, similar to the extended-on-mass-shell scheme~\cite{Fuchs:2003qc}. 
After this procedure, one can safely set the renormalized coupling constants to their physical finite values.

After projecting the $T$-matrix onto the $|LJ\rangle$ partial wave basis, 
one can extract the phase shift, scattering length, and effective range. 
At leading order, since unitarity is exactly satisfied, the phase shift is defined by 
\begin{equation}
	S^{LJ} = e^{2i \delta^{LJ}} = 1+2 i q_{cm} f^{LJ},
\end{equation}
with 
\begin{equation}
	f^{LJ} = -\frac{1}{16\pi^2} \frac{m_N}{E} T^{LJ}.
\end{equation}
Up to NLO, one can obtain the phase shift by performing the perturbative expansion, which gives \cite{Fettes:1998ud}
\begin{equation}
	\delta^{LJ} = \arctan\left(q_{cm} \mathrm{Re}(f^{LJ})\right).
\end{equation}

\section{Results and discussion} \label{sec3}
In this section, we present the phase shifts and effective range expansion for $s$-wave $KN$ scattering up to NLO. 
The masses of pseudoscalar mesons, octet baryons, and vector mesons are taken to be the isospin averaged values: 
$M_K=495.6$ MeV, $m_N=938.9$ MeV, $m_\Lambda=1115.7$ MeV, $m_\Sigma=1193.1$ MeV, $M_{\rho}=775.3$ MeV, $M_{\omega}=782.7$ MeV, and $M_\phi = 1019.5$ MeV. The kaon decay constant is set to $f_K=110.03$ MeV, as adopted from the  Review of Particle Physics~\cite{ParticleDataGroup:2024cfk}. 

\begin{figure*}[t]
\centering
\includegraphics[width=0.9\textwidth]{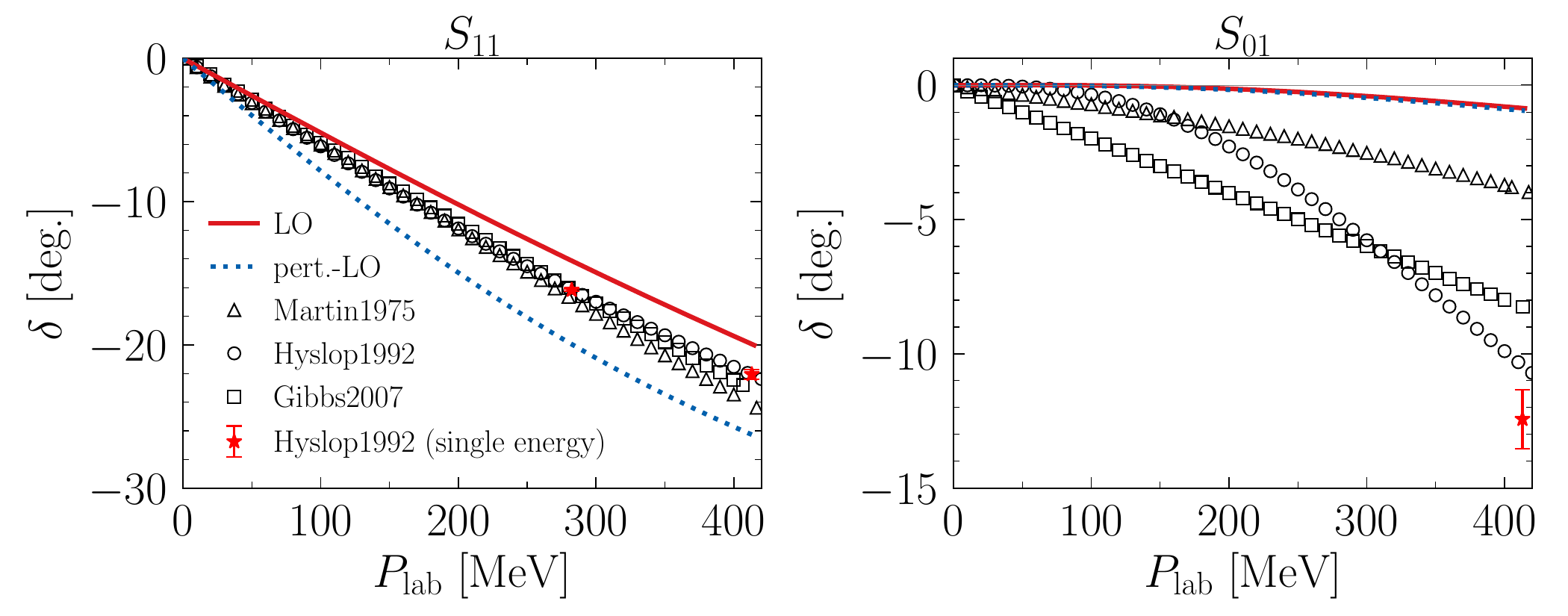} 
\caption{Predictions for the $s$-wave $KN$ scattering phase shifts at LO. The solid red lines denote the non-perturbative LO results with $T$-matrix given as Eq.~(\ref{Eq:TLO}), and the dotted blue lines are the perturbative results at LO with $T=V_\mathrm{LO}$. The triangle, circle, and box datapoints represent the energy-dependent PWAs from Refs.~\cite{Martin:1975gs, Hyslop:1992cs, Gibbs:2006ab}, respectively. The single energy solutions from Ref.~\cite{Hyslop:1992cs} are also presented.}
\label{Fig:KN_LOPS}
\end{figure*}

\subsection{Leading order result}
At leading order, there are no free parameters to be adjusted, allowing us to predict the phase shifts and effective range parameters for the $s$-wave $KN$ scattering. 
In Fig.~\ref{Fig:KN_LOPS}, we show the phase shifts for isospin $I=0$ and $I=1$ channels as a function of the kaon momentum in the laboratory frame, $P_\mathrm{lab}$, up to $400$ MeV, corresponding to the total energy $\sqrt{s}\approx 1525$ MeV. 
We obtain the predictions of phase shifts by iterating the LO potential in the scattering integral equation [Eq.~(\ref{Eq:TLO})], denoted as ``LO'', to investigate the non-perturbative effects in the $KN$ system. In addition, the perturbative results at LO with $T=V_\mathrm{LO}$, denoted as ``pert.-LO'', are also shown in Fig.~\ref{Fig:KN_LOPS}. 
For comparison, the phase shifts from the energy-dependent PWAs of Refs.~\cite{Martin:1975gs, Hyslop:1992cs, Gibbs:2006ab} are also presented.

For the isospin $I=1$ $S_{11}$ channel, since the LO potential is repulsive, the phase shifts monotonically decrease as a function of $P_\mathrm{lab}$. 
Our non-perturbative result at LO is globally close to the empirical PWAs~\cite{Martin:1975gs, Hyslop:1992cs, Gibbs:2006ab} and the single energy solutions of Ref.~\cite{Hyslop:1992cs}, 
and is consistent with the data points up to  $P_\mathrm{lab}=50$ MeV, in comparison with the perturbative ones. 
This indicates that the non-perturbative effect cannot be neglected and can also help to speed up the convergent properties in the SU(3) sector, in comparison with our findings in the SU(2) sector of $\pi N$ scattering~\cite{Ren:2020wid}.  
In contrast, the situation is more complicated in the isospin $I=0$ $S_{01}$ channel. 
The prescriptions of PWAs are not consistent with each other. 
In ChEFT, the LO potential is very weak, and thus the phase shifts from perturbative/non-perturbative treatments are consistent and close to zero. 

Furthermore, we perform the effective range expansion for the $s$-wave scattering via 
\begin{equation}
	p_\mathrm{cm}\, \cot\delta(p_\mathrm{cm}) = \frac{1}{a} + \frac{1}{2} r(p_\mathrm{cm}) \, p_\mathrm{cm}^2 + \mathcal{O}(p_\mathrm{cm}^4),
\end{equation}
with the scattering length $a$ and the effective range $r$, where $p_\mathrm{cm}$ is the relative momentum in the CM frame. For the $S_{11}$ channel, our LO prediction gives the scattering length $a_{KN}^{I=1}=-0.274$ fm using the non-perturbative treatment of the interaction kernel, compared to  $a_{KN}^{I=1}=-0.421$ fm from the perturbative LO calculation. It is clearly seen that the non-perturbative treatment at LO gives a scattering length closer to the experimental range $-0.32\sim -0.28$ fm, as summarized in Ref.~\cite{Dover:1982zh}, than the perturbative approach. 

\begin{figure*}[t]
\centering
\includegraphics[width=0.45\textwidth]{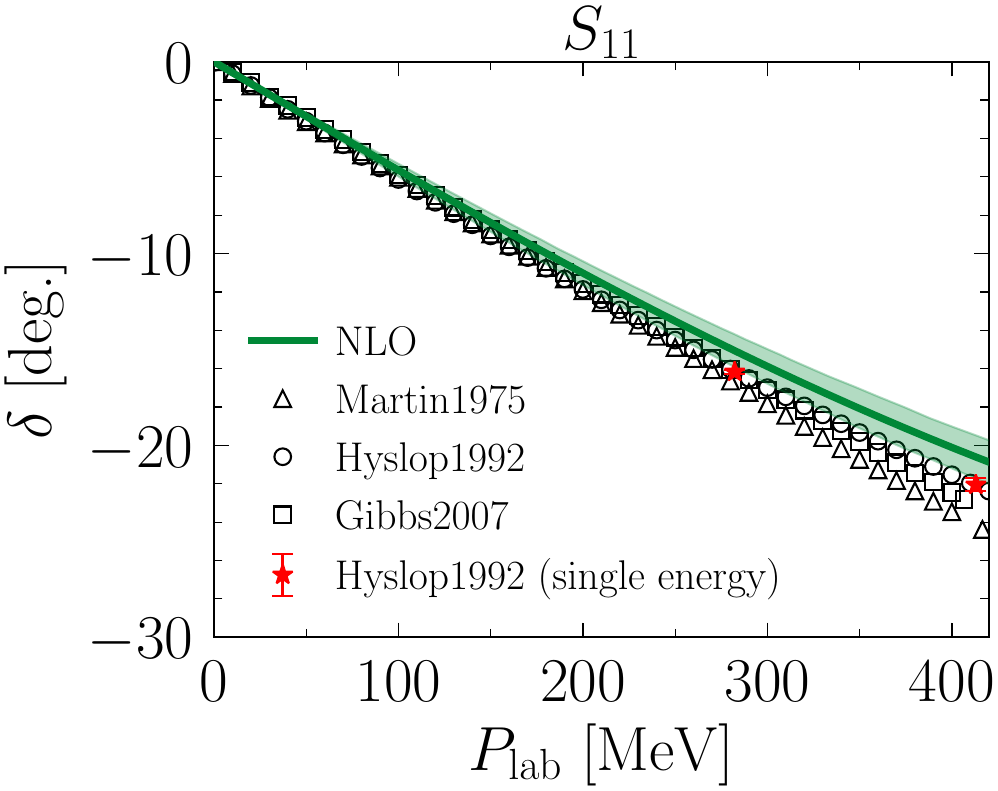} 
~~
\includegraphics[width=0.45\textwidth]{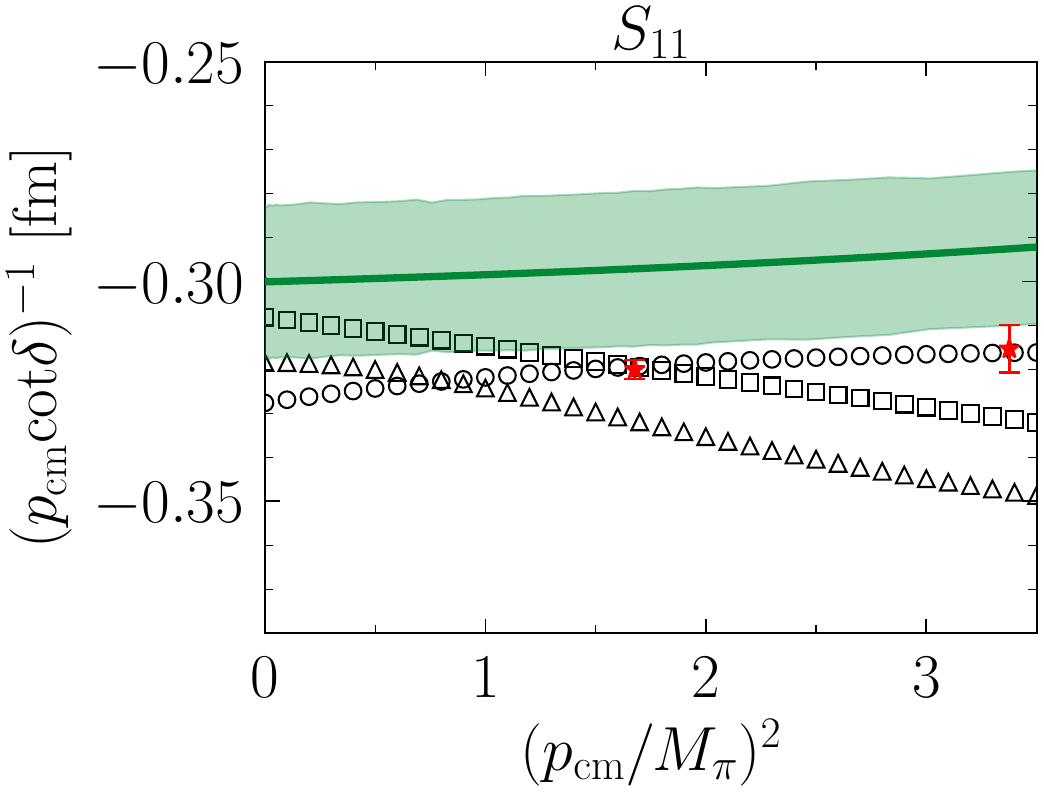} 
\caption{Left panel: description of $KN$ phase shifts in the $S_{11}$ channel up to NLO. Right panel: $\left(p_\mathrm{cm}\cot\delta\right)^{-1}$ as function of $p_\mathrm{cm}/M_\pi$ for the $S_{11}$ $KN$ channel. The solid lines denote the NLO results, and the corresponding light green bands represent the uncertainties at the 68\% confidence level. The narrow band with $p_{cm}=0$ represents the scattering length summarized in Ref.~\cite{Dover:1982zh}. The notations of the datapoints are the same as those given in Fig.~\ref{Fig:KN_LOPS}.}
\label{Fig:KN_S11_NLO}
\end{figure*}

\begin{figure*}[t]
\centering
\includegraphics[width=0.45\textwidth]{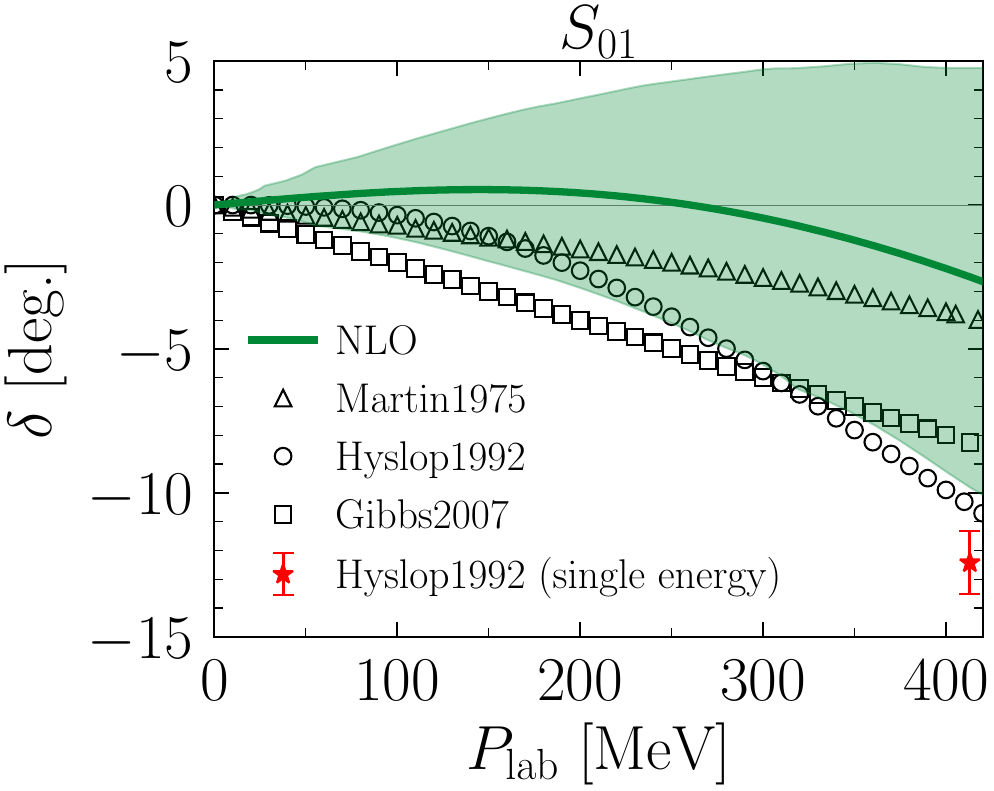} 
~~
\includegraphics[width=0.45\textwidth]{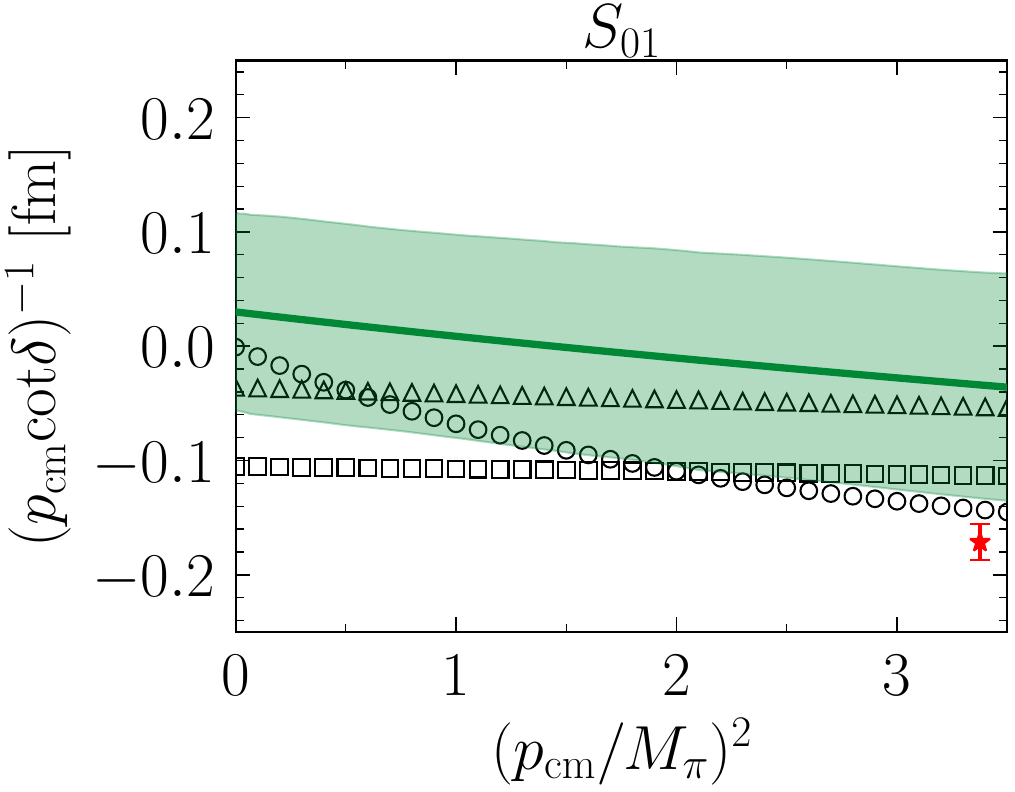} 
\caption{Left panel: description of $KN$ phase shifts in the $S_{01}$ channel up to NLO. Right panel: $\left(p_\mathrm{cm}\cot\delta\right)^{-1}$ as function of $p_\mathrm{cm}/M_\pi$ for the $S_{01}$ $KN$ channel. The solid lines denote the NLO results, and the corresponding light green bands represent the uncertainties at the 68\% confidence level. The narrow band with $p_{cm}=0$ represents the scattering length summarized in Ref.~\cite{Dover:1982zh}. The notations of the datapoints are the same as those given in Fig.~\ref{Fig:KN_LOPS}.}
\label{Fig:KN_S01_NLO}
\end{figure*}

\begin{table}[b]
\caption{Values of parameters $b^{I=0,\,1}$ and $d^{I=0,\,1}$ at NLO in our calculation. The ``*'' denotes parameters with fixed values.}
\label{Tab:LECsValues}
\begin{tabular}{cc|cc}
\hline\hline 
  $b^{I=0}$ [GeV$^{-1}$]  & $d^{I=0}$  [GeV$^{-1}$]  & $b^{I=1}$ [GeV$^{-1}$]  &   $d^{I=1}$  [GeV$^{-1}$]  \\
  \hline 
   $-0.584^*$  & $0.635(759)$  & $-1.924^*$ & $-1.486(360)$   \\
 \hline\hline 
\end{tabular}
\end{table}

\begin{figure*}[t]
\centering
\includegraphics[width=0.9\textwidth]{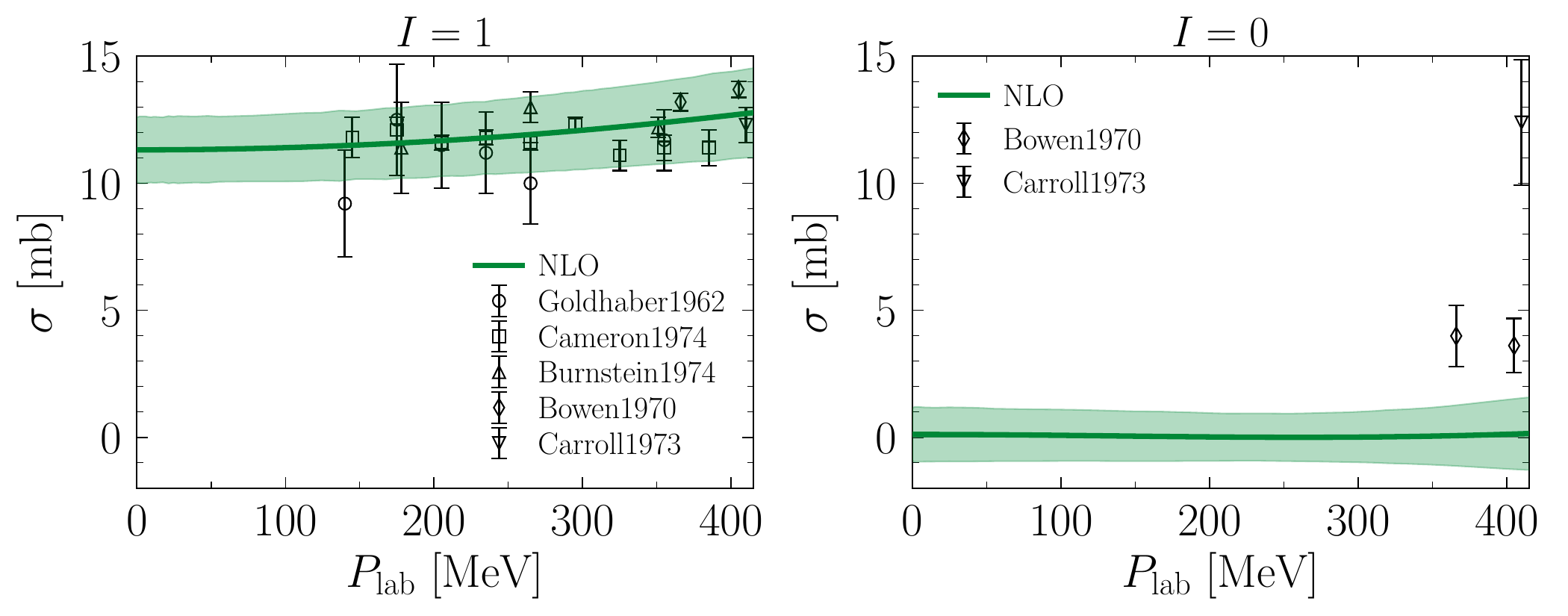} 
\caption{$S$-wave total cross sections for $KN$ scattering at NLO for $I=1$ and $I=0$ sectors. The solid lines denote the NLO results, and the corresponding light green bands represent the uncertainties at the 68\% confidence level. The experimental data for $P_\mathrm{lab}\leq 400$ MeV are taken from Refs.~\cite{Goldhaber:1962zz,Bowen:1970azd,Cameron:1974xx,Burnstein:1974ax}, and the datapoint from Carroll {\it et al.}~\cite{Carroll:1973ux} at $P_\mathrm{lab}=410$ MeV is included for comparison.}
\label{Fig:KN_NLOXSec}
\end{figure*}

\subsection{Next-to-leading order result}
Up to NLO, four unknown parameters $b^{I=0,1}$ and $d^{I=0,1}$ arise from the $\mathcal{O}(p^2)$ contact term, as given in  Eq.~(\ref{Eq:4LECs}). In order to preserve the predictive power of our approach, we fix $b^{I=0,1}$ by adopting the values of $b_0$, $b_D$, and $b_F$ from the RQCD collaboration~\cite{RQCD:2022xux}. 
In that work, octet baryon masses are simulated in lattice QCD with three different quark mass trajectories, providing relatively good constraints on $b_0$, $b_D$, and $b_F$. Applying the Gell-Mann-Okubo mass relation~\cite{Gell-Mann:1962yej,Okubo:1961jc}, one can determine $b_0=-0.389$ GeV$^{-1}$, $b_D=-0.092$ GeV$^{-1}$, $b_F=-0.243$ GeV$^{-1}$ as given in Table 12 of  Ref.~\cite{RQCD:2022xux}, which yields $b^{I=0}=-0.584$ GeV$^{-1}$, $b^{I=1}=-1.924$ GeV$^{-1}$. 

The remaining combinations $d^{I=0}$ and $d^{I=1}$ are determined by reproducing the scattering lengths of the $S_{11}$ and $S_{01}$ partial waves, respectively. As given in Tables 2.2 and 2.3 of Ref.~\cite{Dover:1982zh}, $a_{KN}^{I=1}$ ranges from $-0.33$ fm to $-0.28$ fm, and $a_{KN}^{I=0}$ lies in the range $(-0.11\sim0.18)$ fm, which leads to $a_{KN}^{I=1,\mathrm{exp}}=-0.30\pm 0.03$ fm and $a_{KN}^{I=0,\mathrm{exp}}=0.03\pm 0.15$ fm, as used in Ref.~\cite{Polinder:2005sn}.   
By solving the integrals equations [Eqs.~(\ref{Eq:TLO},\ref{Eq:TNLO})] up to NLO and applying the subtractive renormalization, we are allowed to take the cutoff up to infinity, while in practice we take $\Lambda=20$ GeV to obtain the renormalized $T$-matrix. Then, we reproduce the isospin $I=0,\,1$ scattering lengths at threshold, which leads to $d^{I=0}=0.635(759)$ GeV$^{-1}$ and $d^{I=1}=-1.486(360)$ GeV$^{-1}$. The uncertainties of parameters originate from the uncertainties in scattering lengths~\cite{Dover:1982zh}. For convenience, in Table~\ref{Tab:LECsValues}, we summarize the parameter values used in our NLO calculation. 

After that, we can perform the predictions for the phase shifts and effective range expansion up to NLO, as shown in Figs.~\ref{Fig:KN_S11_NLO} and~\ref{Fig:KN_S01_NLO}. The uncertainty bands are obtained by performing a Monte Carlo sampling of the parameters $d^{I=0,1}$ within their $1\sigma$ uncertainties. The uncertainties of $b^{I=0,1}$ from RQCD are not included, as they are smaller and have a negligible impact on the results.
In the left panels of Figs.~\ref{Fig:KN_S11_NLO} and~\ref{Fig:KN_S01_NLO}, the phase shifts for the $S_{11}$ and $S_{01}$ partial waves are shown up to $P_\mathrm{lab}=400$ MeV. Regarding the $S_{11}$ channel, one can see that the NLO calculation improves the description of phase shifts in comparison with our LO result, 
and is consistent with the empirical PWAs up to $P_\mathrm{lab}\sim 250$ MeV. 
In the right panel of Fig.~\ref{Fig:KN_S11_NLO}, 
we show the results for $1/(p_{cm}\cot\delta)$,  
\begin{equation}
	\frac{1}{p_{cm}\,\cot\delta(p_{cm})} = a - \frac{1}{2}a^2\,r(p_{cm})^2 + \mathcal{O}(p_{cm}^4) 
\end{equation}
as a function of the c.m. momentum $p_{cm}$. In this representation, the intercept denotes the scattering length, and the slope reflects the magnitude of the effective range.
We also present $1/(p_{cm}\cot\delta)$ results from empirical PWAs. In contrast to the phase shifts, $1/(p_{cm}\cot\delta)$ exhibits different evolutionary behavior. The slope of our NLO result is consistent with that of Hyslop et al.~\cite{Hyslop:1992cs}, yielding a negative effective range, which is opposite to the summary result $r\sim (0.32,0.5)$ fm given in Ref.~\cite{Dover:1982zh}. 
Such discrepancy in the effective range reflects that the existing experimental data do not provide sufficient constraints.

Regarding the description of the $S_{01}$ channel in Fig.~\ref{Fig:KN_S01_NLO}, significant  discrepancies exist among different PWAs, as previously noticed, leading to distinct differences in the effective range expansion, as shown in the right panel of Fig.~\ref{Fig:KN_S01_NLO}. Our NLO contribution is very small at low $P_\mathrm{lab}$ in order to reproduce the quasi-zero scattering length of the $S_{01}$ wave. With the increase of $P_\mathrm{lab}$, the NLO correction starts to contribute and generates a repulsive interaction. By taking into account the uncertainty, our NLO description is consistent with the energy-dependence PWAs of Martin~\cite{Martin:1975gs} and Hyslop et al. \cite{Hyslop:1992cs}. 
Nevertheless, the $S_{01}$ wave is found to be particularly sensitive to higher-order contributions, and an improved description is expected once the calculation is extended to NNLO~\cite{Lu:2018zof}.
We also checked that with the increase of total energy up to $\sqrt{s}=1560$ MeV, the phase shift decreases and does not have a sudden increase or a crossing of $90$ degrees. These findings also support the non-existence of the $\Theta^+(1540)$ resonance.

Furthermore, in Fig.~\ref{Fig:KN_NLOXSec}, we present the $s$-wave contribution to the total cross  sections of $KN$ scattering in the isospin $I=0,\,1$ sectors, respectively, via   
\begin{equation}
	\sigma^{LJ}  = \frac{1}{(4\pi)^3} \frac{m_N^2}{s} \left|T^{LJ}\right|^2.
\end{equation}
The experimental data for the total cross section~\cite{Goldhaber:1962zz,Bowen:1970azd,Cameron:1974xx,Burnstein:1974ax,Carroll:1973ux}, which include contributions from all partial waves, are also shown for comparison. In the $I=1$ sector, our NLO result  slowly increases with increasing $P_\mathrm{lab}$, and at threshold the total cross section is $11.3\pm 1.3$ mb.  Our NLO result is consistent with the experimental data, indicating that the $K^+ p\to K^+ p$ cross section (without the Coulomb contribution) is dominated by the $s$-wave contribution. 
In contrast, the $I=0$ total cross section from our $s$-wave calculation is very small, presenting a large deviation from the experimental values. This behavior is consistent with the findings of recent (unitarized) ChPT studies~\cite{Iizawa:2023xsi,Aoki:2018wug}, as well as the most-recent HAL QCD simulations~\cite{Murakami:2025owk}. These results suggest that the $p$-wave contribution plays a dominant role in the $I=0$ total cross section, as also indicated by experimental data analyses~\cite{Stenger:1964vej,Glasser:1977xs,Martin:1975gs}. Therefore, it would be very interesting to extend the present study to include the $p$-wave contribution, especially given the remaining discrepancy in the total cross section between the measurements of Bowen {\it et al.}~\cite{Bowen:1970azd} and Carroll {\it et al.}~\cite{Carroll:1973ux}   in the region of $P_\mathrm{lab}$ around $400$ MeV.

Finally, we employ our NLO formula to investigate the recent lattice QCD simulation on the $s$-wave $KN$ scattering by the HAL QCD collaboration~\cite{Murakami:2025owk}. They obtained a relatively small value of scattering length, $a_{KN}^{I=1}=-0.226(5)\binom{+5}{-0}$ fm, in the $S_{11}$ channel at the physical point with the ``HAL-conf-2023'' configuration~\cite{Aoyama:2024cko}. 
To investigate the effect of the slight differences in meson/baryon masses and the kaon decay constant, relative to the  experimental values, we recalculate the $I=1$ scattering length using the HAL QCD masses and decay constant, and fix the $b^{I=1}$ and $d^{I=1}$ values given in Table~\ref{Tab:LECsValues}. The resulting $a_{KN}^{I=1}$ has a very small deviation from the original value of $-3.0$ fm. Next, we would like to adjust $d^{I=1}$ by reproducing the scattering length given by HAL QCD in the $S_{11}$ channel. An interesting finding is that, in doing so, our predicted effective range $r_{KN}^{I=1}=-0.214$ fm is very close to the HAL QCD value $-0.297(29)\binom{+24}{-0}$ fm.

\section{Summary}\label{sec4}

We have studied $s$-wave $KN$ scattering within a renormalizable ChEFT framework. 
By treating the leading-order interaction nonperturbatively and including higher-order corrections perturbatively in the scattering integral equations, 
we obtain the renormalized scattering $T$-matrix via a subtractive renormalization scheme. 
We extend the renormalizable framework up to next-to-leading order for the $KN$ system. 
After determining the low-energy constants by reproducing the scattering lengths in the  $I=0,\,1$ channels, respectively,  
we present predictions for the phase shifts and effective range expansion up to NLO. 

In the $I=1$ channel, the NLO calculation improves the description of the empirical phase shifts and leads to a negative effective range, 
in agreement with some partial-wave analyses but differing from earlier empirical summaries. 
The $I=0$ interaction at NLO remains very weak with large uncertainties. 
When our results are confronted with the HAL QCD result on the $KN$ scattering, 
an interesting finding is that, if we reproduce the relatively small scattering length obtained by HAL QCD in the $I=1$ channel, 
our predicted effective range is consistent with their determination. 

These results call for more low-energy $KN$ scattering experiments and lattice QCD simulations of the $KN$ system, 
to further constrain the $KN$ interaction and deepen our understanding of strangeness dynamics in QCD. 

Besides, a natural and important extension of the present work would be to carry the calculation of $KN$ scattering up to next-to-next-to-leading order, where the two-pion exchange contributions appear for the first time. Such a study would not only enable a proper order-by-order convergence analysis, but is also expected to improve the description of the $S_{01}$ partial wave. Inspired by findings in the SU(2) sector~\cite{Alarcon:2012kn,Siemens:2016jwj}, where the explicit inclusion of $\Delta(1232)$-isobar contributions was shown to significantly improve the convergence pattern, contributions from the decuplet baryons in the SU(3) sector are similarly expected to be beneficial~(e.g.~\cite{Ren:2013dzt,Xiao:2018rvd} for baryon masses and magnetic moments). These investigations will be delivered in the future works.

\acknowledgements 
This work has been supported in part by Shandong Provincial Natural Science Fund for Excellent Young Scientists Fund Program (Overseas) with project no. 2025HWYQ-015, and by Qilu Youth Scholars Program of Shandong University.

\bibliographystyle{apsrev4-2}
\bibliography{KNrefs}

\end{document}